\documentclass[aps,preprint]{revtex4}%
\usepackage{amsfonts}
\usepackage{amsmath}
\usepackage{amssymb}
\usepackage{graphicx}%
\providecommand{\U}[1]{\protect\rule{.1in}{.1in}}
\begin{document}
\title{Control of the geometric phase and pseudo-spin dynamics on coupled
Bose-Einstein condensates}
\author{E. I. Duzzioni$^{1}$, L. Sanz$^{1\text{,}2}$, S. S. Mizrahi$^{1}$ and M. H. Y.
Moussa$^{3}$}
\affiliation{$^{1}$Departamento de F\'{\i}sica, Universidade Federal de S\~{a}o Carlos,
13565-905, S\~{a}o Carlos, SP, Brazil}
\affiliation{$^{2}$Instituto de F\'{\i}sica, Universidade Federal de Uberl\^{a}ndia, Caixa
Postal 593, 38400-902, Uberl\^{a}ndia, Minas Gerais, Brazil}
\affiliation{$^{3}$Instituto de F\'{\i}sica de S\~{a}o Carlos, Universidade de S\~{a}o
Paulo, Caixa Postal 369, 13560-970, S\~{a}o Carlos, S\~{a}o Paulo, Brazil}

\begin{abstract}
We describe the behavior of two coupled Bose-Einstein condensates in
time-dependent (TD) trap potentials and TD Rabi (or tunneling) frequency,
using the two-mode approach. Starting from Bloch states, we succeed to get
analytical solutions for the TD Schr\"{o}dinger equation and present a
detailed analysis of the relative and geometric phases acquired by the wave
function of the condensates, as well as their population imbalance. We also
establish a connection between the geometric phases and constants of motion
which characterize the dynamic of the system. Besides analyzing the affects of
temporality on condensates that differs by hyperfine degrees of freedom
(internal Josephson effect), we also do present a brief discussion of a one
specie condensate in a double-well potential (external Josephson effect).

\end{abstract}

\maketitle

\section{Introduction}

In recent years, concepts which has been restricted to foundation of quantum
mechanics have been considerably enlarged by spreading out to different
domains of physics. With the introduction of measurements such as separability
\cite{Peres,Horodecki,Simon} and concurrence \cite{Wootters,Werner}, and
the\ wider understanding that entanglement, and so nonlocality, is at the core
of many-body phenomena as quantum-phase transition \cite{Latorre},
superconductivity \cite{BCS} and Bose-Einstein condensation \cite{Pan05}, we
are considerably far from the time when entanglement and nonlocality were
confined to fundamental aspects of quantum mechanics. On the other hand, the
experimental techniques developed over the last decades for manipulating
atom-field interaction have enabled the building of macroscopic atomic
ensembles and the experimental verification of fundamental concepts in
macroscopic scales \cite{Korbicz}.\ It is worth mentioning the rapid growth of
quantum information theory which has conferred to its basic ingredients ---
the phenomena of superposition of states and decoherence, entanglements and
nonlocality --- a great deal of advances towards the accomplishments of
quantum logical devices.

Among the standard tools to generate and detect multipartite entanglements,
experiments in Bose-Einstein condensates (BECs) in dilute gases have deepened
our incursion towards the quantum nature of macroscopic systems. In
particular, experiments with a trapped gas of $^{87}$Rb atoms with two
different hyperfine sublevels prompt the engineering of a Josephson-like
coupling between two condensates by a laser-induced Raman
transition~\cite{Myatt97,Matthews98}. Such \textquotedblleft
internal\textquotedblright\ Josephson effect \cite{LeggettRMP} mediates
intraspecies collisions apart from interspecies one. These atom-atom
interactions empowered the investigation of the dynamics of the relative phase
of coupled condensates \cite{Hall98b} and Rabi oscillations \cite{Matthews99}
on macroscopic systems. Moreover, precise measurements of scattering lengths
has also been accomplished \cite{Hall98a}, aiming to quantify properly the
non-linear dynamics associated with collisions. In the two-mode approximation,
the coupled condensates has been employed to investigate entanglement
dynamics~\cite{Hines03,Sanz03,Pan05} and the possibility to prepare, control
and detect macroscopic superposition
states~\cite{CiracGBEC98,Gordon99,Dunningham01}. Beyond this achievements, the
analysis of the Josephson effect in this two-mode exactly soluble model may
provide a clue for the examination of macroscopic coupling arising in less
tractable form of the general theory of BECs \cite{LeggettRMP}.

Whereas in real experiments the trap potential may be considered to be a
time-independent function, excepting for small fluctuations, time-varying
scattering lengths are usually produced through Feshbach resonances while, as
pointed out in Ref. \cite{LeggettRMP}, the amplitude and phase of the laser
field may vary in time. In this connection, the present paper is devoted to
the TD version of the two-mode Hamiltonian (TMH), where the effective
frequency of the trap potential for both atomic species are TD functions, as
well as the Rabi frequency and the scattering lengths. A similar approach was
employed in Refs.~\cite{Vedral,Chen04} where, however, only the phase of the
external field inducing the Raman transition~was assumed to be a TD slowly
varying function. Instead, our treatment considers time dependence of all
Hamiltonian parameters, focusing on two particular subjects: the analyzes of
geometric phase acquired by wave function of the whole system and the
control\textbf{\ }of the dynamics of pseudo-spin states governed by the TD
TMH. Starting from Bloch states, whose preparation is achieved by applying a
laser pulse~to atoms condensed in a single hyperfine level \cite{Hall98a}, we
demonstrate that its evolution, visualized as a vector on the Bloch sphere,
can be used to control the geometric phase and the population imbalance
following from the whole wave function of the condensates. Our treatment also
permits a detailed analyzes of the relative phase between the condensed states.

In Ref.\cite{Milburn}, the authors studied the dynamics of a strongly driven
two-coupled BECs in two spatially localized modes of a double-well potential,
where the tunneling coupling between the two modes is periodically modulated.
In our work we also study the TD TMH associated to the \textquotedblleft
external\textquotedblright\ Josephson effect, analyzing its differences with
relation of the \textquotedblleft internal\textquotedblright\ Josephson effect.

Similarly to the above mentioned fundamental phenomena, the geometric phase
has overtaken its striking rule on fundamental physics to widening our
understanding of phenomena as quantum Hall effect \cite{Baily,Bruno,Kats} and
for the implementation of fault-tolerant quantum gates \cite{Zanardi}. After
its discovery by Berry on adiabatic processes \cite{Berry98}, it has been
generalized to nonadiabatic~\cite{Aharonovl87}, noncyclic~\cite{Samuel88} and
nonunitary~\cite{Tong04,Piza} quantum evolutions. Recently, it has been
investigated in different areas of physics, ranging from BECs \cite{Vedral}
and cavity quantum electrodynamics \cite{Carollo,Duzzioni} to condensed matter
\cite{Bliokh} and quantum information theory \cite{Zanardi}. In particular,
the Berry phase of mesoscopic spin in Bose-Einstein condensates, induced by a
TD slowly varying driven field, has been investigated under the TMH
\cite{Vedral,Chen04}. In our treatment, the evolution of the geometric phase
of this mesoscopic system is evaluated in a more general scenario, where all
the Hamiltonian parameters are assumed to be TD.

The paper is organized as follows. In Sec.~II we introduce and solve the
Schr\"{o}dinger equation associated to the TD TMH, presenting the evolution
operator. The dynamics of BECs for initial Bloch states is analyzed in Sec.
III, where we show that they remain as Bloch states apart from a global phase
factor accounting for the elastic collision terms. The geometric phase
acquired by the state vector of the system is presented in Sec. IV and a
detailed analyzes of its time evolution is found in Sec. V for different
regimes of the parameters. In Sec. VI we take the problem of a TMH from a
different perspective, considering the external Josephson effect instead of
the internal one. Finally, Sec. VII is devoted to our concluding remarks.

\section{The time-dependent TMH}

Under the two-mode approximation, where the quantum field operators $\Psi
_{a}=\varphi_{a}\left(  \mathbf{r},t\right)  a$ and $\Psi_{b}=\varphi
_{b}(\mathbf{r},t)b$ are restricted to the fundamental states $\varphi_{\ell
}\left(  \mathbf{r},t\right)  $ ($\ell=a,b$) \cite{CiracGBEC98,Milburn1997},
the coupled Bose-Einstein condensates are described by the TD Hamiltonian
($\hbar=1$)
\begin{align}
H\left(  t\right)   &  =\sum_{\ell=a,b}\left[  \omega_{\ell}\left(  t\right)
\ell^{\dagger}\ell+\gamma_{\ell}\left(  t\right)  \ell^{\dagger}\ell^{\dagger
}\ell\ell\right]  +\gamma_{ab}\left(  t\right)  a^{\dagger}ab^{\dagger
}b\nonumber\\
&  -g\left(  t\right)  \left(  e^{-i\delta\left(  t\right)  }a^{\dag
}b+e^{i\delta\left(  t\right)  }ab^{\dag}\right)  \text{,} \label{1}%
\end{align}
where $a$ and $b$ are standard bosonic annihilation operators, associated with
condensation in hyperfine levels $\left\vert 2,1\right\rangle $ and
$\left\vert 1,-1\right\rangle $, respectively
\cite{CiracGBEC98,Gordon99,Villain99}. The phase $\delta\left(  t\right)  $ is
associated to the detuning $\Delta(t)$ from the Raman resonance between the
atomic transition $\left\vert 2,1\right\rangle $ $\leftrightarrow\left\vert
1,-1\right\rangle $, which may be a TD function (by varying the laser
frequency), through the expression $\delta\left(  t\right)  =\int_{t_{0}}%
^{t}\Delta(\tau)$ $d\tau+$ $\delta_{0}$. The TD trap frequencies $\omega
_{\ell}$, the interspecies and intraspecies collision parameters $\gamma_{ab}$
and $\gamma_{\ell}$, and the Rabi frequency $g$, follow from
\begin{subequations}
\label{2}%
\begin{align}
\omega_{\ell}\left(  t\right)   &  =\int d^{3}{\mathbf{r}}\varphi_{\ell}%
^{\ast}\left(  \mathbf{r},t\right)  \left[  -\frac{1}{2m}\nabla^{2}+V_{\ell
}\left(  \mathbf{r},t\right)  \right]  \varphi_{\ell}\left(  \mathbf{r}%
,t\right)  \text{,}\label{2a}\\
\gamma_{\ell}\left(  t\right)   &  =\frac{4\pi A_{\ell}\left(  t\right)  }%
{2m}\int d^{3}{\mathbf{r}}\left\vert \varphi_{\ell}\left(  \mathbf{r}%
,t\right)  \right\vert ^{4}\text{,}\label{2f}\\
\gamma_{ab}\left(  t\right)   &  =\frac{4\pi A_{ab}\left(  t\right)  }{m}\int
d^{3}{\mathbf{r}}\left\vert \varphi_{a}\left(  \mathbf{r},t\right)
\varphi_{b}\left(  \mathbf{r},t\right)  \right\vert ^{2}\text{,}\label{2g}\\
g\left(  t\right)   &  =\frac{\Omega\left(  t\right)  }{2}\int d^{3}%
{\mathbf{r}}\varphi_{a}^{\ast}\left(  \mathbf{r},t\right)  \varphi_{b}\left(
\mathbf{r},t\right)  , \label{2d}%
\end{align}
where $m$ is the atomic mass. We assume that the time dependence of the trap
potential $V_{\ell}\left(  \mathbf{r},t\right)  $ is generated by
adiabatically varying the trapping magnetic field.\textbf{\ }Such adiabatic
variation of the trapping field has been assumed to ensure the validity of the
two-mode approximation. The time-varying scattering lengths $A_{ab}(t)$ and
$A_{\ell}(t)$, are accomplished via Feshbach resonances, by tuning a bias
magnetic field \cite{Vogels}. Finally, as mentioned above, in real experiments
with atomic BECs the Rabi frequency may be a time-varying function since the
amplitude and phase of the pumping fields may vary in time \cite{LeggettRMP}.

Except for the Josephson-like coupling, the Fock states are eigenstates of all
the terms in Hamiltonian (\ref{1}). Thus, in order to get rid of this TD
coupling in (\ref{1}), we consider a transformation with the unitary operator%

\end{subequations}
\begin{equation}
V(t)=\exp\left[  \frac{r(t)}{2}\left(  \operatorname*{e}\nolimits^{i\phi
(t)}ab^{\dagger}-\operatorname*{e}\nolimits^{-i\phi(t)}a^{\dagger}b\right)
\right]  \text{,} \label{3}%
\end{equation}
(analogous to that defined in Ref.~\cite{Chen04}) to obtain the transformed Hamiltonian%

\begin{equation}
\mathcal{H}(t)=V^{\dagger}HV-iV^{\dagger}\partial_{t}V=\sum_{\ell
=a,b}\widetilde{\omega}_{\ell}(t)n_{\ell}+\mathcal{H}_{el}(t)+\mathcal{H}%
_{inel}(t)\text{,} \label{4}%
\end{equation}
where $n_{\ell}=\ell^{\dagger}\ell$ is the number operator associated to each
condensate having effective frequency
\begin{equation}
\widetilde{\omega}_{\ell}\left(  t\right)  =\omega_{\ell}\left(  t\right)
+\left(  2\delta_{\ell b}-1\right)  g\left(  t\right)  \cos\left[  \phi\left(
t\right)  -\delta\left(  t\right)  \right]  \tan\left[  r\left(  t\right)
/2\right]  \text{.} \label{5}%
\end{equation}
In the framework associated to the transformation (\ref{3}), the system ends
up with an inelastic collision term apart from the elastic one already present
in (\ref{1}). The Hamiltonians accounting for such interactions, also weighted
by the TD function $\Lambda(t)=\gamma_{a}(t)+\gamma_{b}(t)-\gamma_{ab}(t)$,
are given by%

\begin{subequations}
\label{6}%
\begin{align}
\mathcal{H}_{el}(t)  &  =\left\{  \gamma_{a}(t)\cos^{2}\left[  r(t)/2\right]
+\gamma_{b}(t)\sin^{2}\left[  r(t)/2\right]  -\frac{\Lambda(t)}{4}\sin
^{2}\left[  r(t)\right]  \right\}  \left(  a^{\dagger}\right)  ^{2}%
a^{2}\nonumber\\
&  +\left\{  \gamma_{a}(t)\sin^{2}\left[  r(t)/2\right]  +\gamma_{b}%
(t)\cos^{2}\left[  r(t)/2\right]  -\frac{\Lambda(t)}{4}\sin^{2}\left[
r(t)\right]  \right\}  \left(  b^{\dagger}\right)  ^{2}b^{2}\nonumber\\
&  +\left\{  \gamma_{ab}(t)+\Lambda(t)\sin^{2}\left[  r(t)\right]  \right\}
a^{\dagger}ab^{\dagger}b\text{,}\label{6a}\\
\mathcal{H}_{inel}(t)  &  =\left\{  \frac{\left[  \gamma_{b}(t)-\gamma
_{a}(t)\right]  }{2}\sin\left[  r(t)\right]  -\frac{\Lambda(t)}{4}\sin\left[
2r(t)\right]  \right\}  \operatorname*{e}\nolimits^{-i\phi(t)}\left(
a^{\dagger}\right)  ^{2}ab\nonumber\\
&  +\left\{  \frac{\left[  \gamma_{b}(t)-\gamma_{a}(t)\right]  }{2}\sin\left[
r(t)\right]  +\frac{\Lambda(t)}{4}\sin\left[  2r(t)\right]  \right\}
\operatorname*{e}\nolimits^{-i\phi(t)}a^{\dagger}b^{\dag}b^{2}\nonumber\\
&  +\frac{\Lambda(t)}{4}\sin^{2}\left[  r(t)\right]  \left(  \operatorname*{e}%
\nolimits^{-i\phi(t)}a^{\dagger}b\right)  ^{2}+\mathrm{h{.c.}}\text{.}
\label{6b}%
\end{align}
The form of Hamiltonian (\ref{4}) is established provided that the TD
parameters $r(t)$ and $\phi(t)$ satisfy the coupled differential equations
\end{subequations}
\begin{subequations}
\label{7}%
\begin{align}
\overset{.}{r}(t)  &  =2g(t)\sin\left[  \phi\left(  t\right)  -\delta\left(
t\right)  \right]  \mathrm{,}\label{7a}\\
\overset{.}{\phi}(t)  &  =\omega(t)+2g(t)\cot\left[  r(t)\right]  \cos\left[
\phi\left(  t\right)  -\delta\left(  t\right)  \right]  \text{{,}} \label{7b}%
\end{align}
where
\end{subequations}
\begin{equation}
\omega\left(  t\right)  =\omega_{a}\left(  t\right)  -\omega_{b}\left(
t\right)  \text{.} \label{8}%
\end{equation}
The expression (\ref{8}) represents an effective frequency for the system
composed by the two-mode condensate, which plays an important role in the
solutions of the characteristic equations (\ref{7}). In the Appendix we
present a comprehensive analyzes of the analytical solutions of the Eqs.
(\ref{7}) for the on- and off-resonant regimes which are defined by comparing
the effective frequency $\omega(t)$ with the detuning between the laser field
and Raman transition $\Delta(t)$. The on-resonant regime, where $\Delta
(t)=\omega\left(  t\right)  $, implies that the detuning from Raman transition
must equals the effective frequency of the two-mode condensate. Otherwise, we
have the off-resonant regime, where $\Delta(t)=\omega(t)-\varpi$, $\varpi$
being some constant.

Similar coupled differential equations were obtained by Smerzi \textit{et al}.
\cite{Smerzi 1997} in a semi-classical treatment of the double-well problem,
and by Chen \textit{et al}. \cite{Chen04} in a full quantum approach of the
BECs in the two-mode approximation. In both references all the parameters in
their Hamiltonians are constants, except for $\delta\left(  t\right)  $ which,
in Ref. \cite{Chen04}, is an adiabatically time-varying parameter.

After the experiments by Hall and co-workers with $^{87}$Rb \cite{Hall98a},
where the scattering lengths satisfy the relation $A_{a}:A_{ab}:A_{b}%
=1.03:1:0.97$, and consequently $\gamma_{a}\simeq\gamma_{b}\simeq\gamma
_{ab}/2$ (assuming spatial Gaussian function), a number of papers have driven
attention to this particular case \cite{Liliana,Chen04} whose Schr\"{o}dinger
equation is exactly soluble. However, if the proportion $1.03:1:0.97$ is
broken, the system is not exactly integrable, but admits approximated
solutions as shown in Refs. \cite{Chen04,Vedral}. In our paper we are
concerned with specific solutions of the characteristic equations (\ref{7}),
under the rotating-wave approximation, which turn negligible the contribution
of the inelastic interactions compared to the elastic one. This is done by
substituting the solutions for $r(t)$ and $\phi(t)$, obtained in the Appendix,
into Eqs. (\ref{5}) and (\ref{6}), and rewriting Hamiltonian (\ref{4}) in the
interaction picture. Thus, after a time average of the TD parameters appearing
in this Hamiltonian, we analyze the conditions leading to the effective
interaction
\begin{equation}
\mathcal{H}_{eff}(t)\simeq\sum\nolimits_{\ell=a,b}\widetilde{\omega}_{\ell
}(\tau)n_{\ell}+\mathcal{H}_{el}(\tau)\text{.} \label{9}%
\end{equation}
By adopting this procedure, where the inelastic interactions becomes
despicable, we get he evolution operator $\mathcal{U}\left(  t,t_{0}\right)
=\exp\left(  -i\int_{t_{0}}^{t}\mathcal{H}_{eff}(\tau)d\tau\right)  $ and,
consequently, a prepared state $\left\vert \psi\left(  t_{0}\right)
\right\rangle $ evolves according to Hamiltonian (\ref{1}) as
\begin{equation}
\left\vert \psi\left(  t\right)  \right\rangle =V\left(  t\right)
\mathcal{U}\left(  t,t_{0}\right)  V^{\dagger}\left(  t_{0}\right)  \left\vert
\psi\left(  t_{0}\right)  \right\rangle \text{.} \label{10}%
\end{equation}

\section{Dynamics of BECs for initial Bloch states}

Following Arecchi \textit{et al.}~\cite{Arecchi72} and Dowling \textit{et
al.}~\cite{Dowling94}, we recall that the Bloch states (BS), also called
atomic coherent states, spanned in the Dicke basis $\left\vert
j,m\right\rangle $, where $j=N/2$ and $\left\vert m\right\vert \leq j$, are
obtained through a specific rotation on the reference state $\left\vert
j,j\right\rangle $,
\begin{equation}
\left\vert \alpha,\beta\right\rangle =e^{\frac{\alpha}{2}\left(  e^{i\beta
}\hat{J}_{-}-e^{-i\beta}\hat{J}_{+}\right)  }\left\vert j,j\right\rangle
\text{,} \label{15}%
\end{equation}
where $N$ is the number of condensed particles, $J_{+}$, $J_{-}$ (together
with $J_{z}$), are the generators of the $su(2)$ algebra, and $\left(
\alpha,\beta\right)  $ are the polar and the azimuthal angles, respectively,
defined on the Bloch sphere. The Heisenberg angular-momentum uncertainty
relation for the BS reduces to
\begin{equation}
\langle\left(  \Delta J_{x^{\prime}}\right)  ^{2}\rangle\langle\left(  \Delta
J_{y^{\prime}}\right)  ^{2}\rangle=\frac{1}{4}\left\vert \langle J_{z^{\prime
}}\rangle\right\vert ^{2}\text{,} \label{16}%
\end{equation}
with the mean values being calculated in a rotated coordinate system
$x^{\prime},y^{\prime},z^{\prime}$,where $z^{\prime}$ is an axis in the
($\alpha$,$\beta$) direction through the center of the Bloch sphere.
Therefore, the Bloch vector is defined as the unit vector
\begin{equation}
\mathbf{n}=\left(  \sin{\alpha}\cos{\beta}\text{,}\sin{\alpha}\sin{\beta
}\text{,}\cos{\alpha}\right)  \text{,} \label{17}%
\end{equation}
in $z^{\prime}$-axis. Using the Schwinger relations
\begin{subequations}
\label{17l}%
\begin{align}
J_{x}  &  =\frac{1}{2}\left(  a^{\dagger}b+ab^{\dagger}\right)  \text{,}%
\label{17la}\\
J_{y}  &  =\frac{1}{2i}\left(  a^{\dagger}b-ab^{\dagger}\right)
\text{,}\label{l7lb}\\
J_{z}  &  =\frac{1}{2}\left(  a^{\dagger}a-b^{\dagger}b\right)  \text{.}
\label{17lc}%
\end{align}
with $J_{\pm}=J_{x}\pm iJ_{y}$ and the basis states $\left\{  \left\vert
j,m\right\rangle =\left\vert N/2,\left(  N_{a}-N_{b}\right)  /2\right\rangle
\right\}  \equiv\left\{  \left\vert N_{a}\right\rangle \left\vert
N_{b}\right\rangle \equiv\left\vert N_{a},N_{b}\right\rangle \right\}  $,
where $N_{a}$ and $N_{b}$ ($N=N_{a}+N_{b}$) stand for the number of atoms in
the condensates, such that $\left\vert j,j\right\rangle \Longleftrightarrow
\left\vert N,0\right\rangle $, it is straightforward to check that the BS can
be defined through bosonic operators as
\end{subequations}
\begin{equation}
\left\vert \alpha,\beta\right\rangle =\frac{1}{\sqrt{N!}}\left[  \cos\left(
\frac{\alpha}{2}\right)  a^{\dagger}+\sin\left(  \frac{\alpha}{2}\right)
\operatorname*{e}\nolimits^{i\beta}b^{\dagger}\right]  ^{N}\left\vert
0,0\right\rangle \text{.} \label{18}%
\end{equation}
This state has a well-defined relative phase $\beta$ between the two bosonic modes.

Now, it is evident from relations (\ref{17l}) that the unitary transformation
$V\left(  t\right)  $ turns to be exactly the rotation operator $e^{\frac
{\alpha}{2}\left(  e^{i\beta}\hat{J}_{-}-e^{-i\beta}\hat{J}_{+}\right)  }$, if
one considers $r\left(  t\right)  =\alpha$ and $\phi\left(  t\right)  =\beta$.
Therefore, it is easy to demonstrate through Eq. (\ref{10}) that an initial BS
$\left\vert \psi\left(  t_{0}\right)  \right\rangle =$ $\left\vert \alpha
_{0},\beta_{0}\right\rangle $ $=$ $\left\vert r_{0},\phi_{0}\right\rangle $
evolves to another BS%

\begin{align}
\left\vert {\psi}(t)\right\rangle  &  =\frac{e^{-iN\varphi_{N}(t)}}{\sqrt{N!}%
}\left[  \cos\left(  \frac{r(t)}{2}\right)  a^{\dagger}+\sin\left(
\frac{r(t)}{2}\right)  e^{i\phi\left(  t\right)  }b^{\dagger}\right]
^{N}\left\vert 0,0\right\rangle \text{,}\nonumber\\
&  =e^{-iN\varphi_{N}(t)}\left\vert r(t),\phi\left(  t\right)  \right\rangle
\label{19}%
\end{align}
apart from the global phase factor $e^{-iN\varphi_{N}(t)}$, where%
\begin{align}
\varphi_{N}(t)  &  =\int_{t_{0}}^{t}\left\{  \widetilde{\omega}_{a}%
(\tau)+(N-1)\left[  \gamma_{a}(\tau)\cos^{2}\left[  r(\tau)/2\right]  \right.
\right. \nonumber\\
&  +\left.  \left.  \gamma_{b}(\tau)\sin^{2}\left[  r(\tau)/2\right]
-\frac{\Lambda(\tau)}{4}\sin^{2}\left[  r(\tau)\right]  \right]  \right\}
d\tau\text{.} \label{19a}%
\end{align}
The relative phase between the condensates $\phi(t)$ is associated to the mean
values $\left\langle J_{x}(t)\right\rangle =N\sin\left[  r(t)\right]
\cos\left[  \phi(t)\right]  /2$ and $\left\langle J_{y}(t)\right\rangle
=N\sin\left[  r(t)\right]  \sin\left[  \phi(t)\right]  /2$, whereas $r(t)$ is
related to the population imbalance, $\Delta N(t)=\left\langle N_{a}%
-N_{b}\right\rangle =2\left\langle J_{z}(t)\right\rangle $, through the
relation
\begin{equation}
\Delta N(t)=N\cos\left[  r(t)\right]  \text{.} \label{20}%
\end{equation}

We stress that according to the evolution operator, Eq. (\ref{10}), the
evolved state (\ref{19}) remains a BS (apart from a global phase factor),
since $\alpha_{0}=r_{0}$ and $\beta_{0}=\phi_{0}$. Note that the collision
parameters are restricted to the global phase $\operatorname{e}^{-iN\varphi
_{N}(t)}$, being irrelevant to the analyzes developed below of the
nonadiabatic geometric phases acquired by the state vector. On the other hand,
collisions become relevant when considering other initial states as the
product of Glauber's coherent states $\left\vert \alpha_{0}\right\rangle
\left\vert \beta_{0}\right\rangle $ instead of the BS \cite{Liliana}.

\section{Geometric phases of the BS}

To study the evolution of the geometric phase in the two-mode BECs we use the
kinematic approach developed by Mukunda and Simon \cite{Mukunda1993}, where
the geometric phase $\phi_{G}$ is obtained as the difference between the total
phase $\phi_{T}(t)=\arg(\left\langle \psi(t_{0})\right\vert \left.
\psi(t)\right\rangle )$ and the dynamical phase $\phi_{D}(t)=-i\int_{t_{0}%
}^{t}\left\langle \psi(\tau)\right\vert \frac{\partial}{\partial\tau
}\left\vert \psi(\tau)\right\rangle d\tau$, resulting in
\begin{equation}
\phi_{G}(t)=\arg(\left\langle \psi(t_{0})\right\vert \left.  \psi
(t)\right\rangle )+i\int_{t_{0}}^{t}\left\langle \psi(\tau)\right\vert
\frac{\partial}{\partial\tau}\left\vert \psi(\tau)\right\rangle d\tau.
\label{23}%
\end{equation}
The expressions for $\phi_{G}(t)$ and $\phi_{D}(t)$ as function of the system
parameters are, respectively,%
\begin{align}
\phi_{G}(t)  &  =N\arg\{\cos\left(  r_{0}/2\right)  \cos\left[  r(t)/2\right]
+\operatorname{e}^{i\left[  \phi(t)-\phi_{0}\right]  }\sin\left(
r_{0}/2\right)  \sin\left[  r(t)/2\right]  \}\nonumber\\
&  -\frac{N}{2}\int_{t_{0}}^{t}\overset{.}{\phi}(\tau)\left\{  1-\cos\left[
r(\tau)\right]  \right\}  d\tau, \label{24}%
\end{align}
and%
\begin{equation}
\phi_{D}(t)=-N\left[  \varphi_{N}(t)-\varphi_{N}(t_{0})\right]  +\frac{N}%
{2}\int_{t_{0}}^{t}\overset{.}{\phi}(\tau)\left\{  1-\cos\left[
r(\tau)\right]  \right\}  d\tau, \label{25}%
\end{equation}
where $\varphi_{N}(t)$ was given in Eq. (\ref{19a}). In
Ref.\cite{Balakrishnan}, the authors also obtain Eq.(\ref{24}), through the
Gross-Pitaevskii equation, for the geometric phase acquired by the wave
function of a BEC in the double-well problem, under the two-mode approximation.

The time evolution of the BS can be followed\textbf{\ }on the Bloch sphere
through the vector $\mathbf{n}\left(  t\right)  =(\sin[r(t)]\cos[\phi(t)]$,
$\sin[r(t)]\sin[\phi(t)]$, $\cos[r(t)])$, for the different solutions $r(t)$
and $\phi(t)$ presented in the Appendix. To follow such evolution and,
consequently, to analyze its geometric phase $\phi_{G}(t)$, we must estimate
the integrals in Eq.(\ref{2}), i.e., the typical values for trap frequencies,
Josephson-like coupling, intraspecies, and interspecies collision rates. For
the sake of simplicity we model the effective frequency for both atomic
species as harmonic trap potentials where the TD distribution for each
condensate $\varphi_{\ell}\left(  \mathbf{r},t\right)  $ can be approximated
by a stationary Gaussian function such that
\begin{equation}
\varphi_{\ell}\left(  \mathbf{r}\right)  =\left(  \frac{1}{2\pi x_{\ell}^{2}%
}\right)  ^{3/4}e^{-\mathbf{r}^{2}/4x_{\ell}^{2}}\text{,} \label{26}%
\end{equation}
where $x_{\ell}=\sqrt{\hbar/2m\omega_{\ell}}$ stands for the position
uncertainty in each harmonic oscillator ground state \cite{Milburn}. With this
assumption the integrals in Eq.(\ref{2}) are immediately estimated using
typical physical parameters of the experiments with $^{87}$Rb atoms
\cite{Hall98a,Albiez2005,Gordon1998}: $m=1.4\times10^{-25}$ Kg, $\omega_{\ell
}\sim10^{1-2}$ Hz, $A_{\ell}\sim5$ nm, and $\Omega\sim10^{3}$ Hz\textbf{.} To
obtain some insights of \ the pseudo-spin dynamics under the TD Hamiltonian
parameters, we consider the trap and Rabi frequencies as harmonic functions,
oscillating around the typical constant values, as follow
\begin{subequations}
\label{27}%
\begin{align}
\omega_{\ell}\left(  t\right)   &  =\omega_{\ell0}+\widetilde{\omega}_{\ell
}\sin{\left(  \chi_{\ell}t+\xi_{\ell}\right)  ,}\label{27a}\\
g(t)  &  =g_{0}+\widetilde{g}\sin(\mu t)\text{,} \label{27b}%
\end{align}
with the parameters $\omega_{\ell0}$, $\widetilde{\omega}_{\ell}$, $\chi
_{\ell}$, $\xi_{\ell}$, $g_{0}$, $\widetilde{g}$, and $\mu$ being constant.
Since the elastic collisions contribute only to a global phase factor, they
will be assumed as the standard constant parameters in the literature
\end{subequations}
\begin{subequations}
\label{28}%
\begin{align}
\gamma_{\ell}  &  =\frac{4\pi A_{\ell}}{2m}\text{,}\label{28a}\\
\gamma_{ab}  &  =\frac{4\pi A_{ab}}{m}\text{.} \label{28b}%
\end{align}

\section{Nonadiabatic Geometric Phases and Pseudo-Spin Dynamics in BECs}

In this section we present a detailed study of the geometric phase acquired by
the whole wave function of the two-mode condensates. To this end, we plot the
time evolution of the geometric phase (\ref{24})\textbf{, }and analyze its
behavior through the evolution of Bloch vector (a map of the wave function of
the BECs on Bloch sphere). This procedure allows for a better understanding
and visualization of the concept of geometric phase for open trajectories
introduced by Samuel and Bhandari \cite{Samuel88}. In particular, we are
interested in the dependence of the geometric phase on the constant of motions
coming from the solutions of the characteristic equations (\ref{7}) and also
on the time dependence of the Hamiltonian parameters. In spite of the general
solutions presented in the Appendix for these equations, we have assumed, for
the analysis developed below, $N=1$, $\delta_{0}=0$, and $g_{0}=625\pi$ Hz.

\subsection{Solutions for $r$ constant}

Before analyzing the geometric phases for the on- and off-resonant solutions,
it is instructive to present their evolutions for the simple case where the
parameter $r$ is kept constant while $\phi\left(  t\right)  $ obeys
Eq.(\ref{A9}) (since $g=0$). In this case, the expression for the geometric
phase coming from Eq. (\ref{24}), becomes%

\end{subequations}
\begin{equation}
\phi_{G}(t)=N\left\{  \arg\left\{  \cos^{2}\left(  r/2\right)  +e^{-i\left[
\phi(t)-\phi_{0}\right]  }\sin^{2}\left(  r/2\right)  \right\}  -\frac{\left(
1-\cos r\right)  }{2}\left[  \phi(t)-\phi_{0}\right]  \right\}  \text{.}
\label{fg1}%
\end{equation}

In Fig.1 the absolute value for $\phi_{G}(t)$ is plotted against the
dimensionless $\tau=\omega_{a0}t$, assuming typical values $\omega
_{a0}=2\omega_{b0}=62.5\pi$ Hz. For $r=\pi/2$, with the Bloch vector standing
on the equatorial plane, and $\widetilde{\omega}_{a}=\widetilde{\omega}_{b}%
=0$, the geometric phase evolves by jumps, as indicated by the thick solid
line. These jumps occur every time the relative phase $\phi$ connecting the
final to the initial Bloch vectors equals $\left(  2n+1\right)  \pi$, $n$
being an integer. The jump discontinuities occur because there are an infinite
number of small geodesic-lengths connecting the vectors extremities, rendering
the geometric phase undefined \cite{Polavieja}. On the other hand, we observe
that before jumping to $\phi_{G}(\tau)=\pi$, i.e., for $\tau<2\pi$, the
geometric phase remain null since the small geodesic-length connecting the
extremities equals the Bloch-vector trajectory itself. As soon as the Bloch
vector acquires a relative phase larger than $\pi$, the small geodesic-length
connecting the extremities completes a loop over the equator, making the
acquired geometric phase proportional to $nN\pi$, where $n$, as defined above,
turns out to be the winding number, i.e., the number of loops around the
$z$-axis of the sphere.

The same interpretation given above for the geometric phase holds for the two
other curves obtained for $r=\pi/2.1$, except that the jump discontinuities
are substituted by high-slope curves around the points where $\phi=(2n+1)\pi$.
Moreover, the net effect coming from the TD parameters of the Hamiltonian is
to delay or advance the sequential increments of the relative phase $\phi$
and, consequently, of the geometric phase, as observed from Fig.1. The solid
line, obtained for $r=\pi/2.1$ and $\widetilde{\omega}_{a}=\widetilde{\omega
}_{b}=0$, shows that the increments of the geometric phase, besides being
smaller, present the same rate of variation when compared to the case
$r=\pi/2$. When the trap frequencies are oscillating functions, with
$\widetilde{\omega}_{a}=\widetilde{\omega}_{b}=\omega_{a0}/4$, $\chi_{a}%
=\chi_{b}=\omega_{a0}/2$, $\xi_{a}=0$, and $\xi_{b}=\pi/2$, the
time-dependence shown by the dashed line, results in the oscillations of the
time intervals between the increments of the geometric phase.

To better visualize the above discussion about geometric phases acquired in
open trajectories, in Fig.2 we plot the evolution of the Bloch vectors for the
cases $r=\pi/2$ and $r=\pi/2.1$, with $\widetilde{\omega}_{a}=\widetilde
{\omega}_{b}=0$, considering the same time interval $\tau=4\pi$. The black and
grey vectors indicate the coincident positions of the initial and final Bloch
vectors, for the cases $r=\pi/2.1$ and $r=\pi/2$, respectively, after a
complete rotation around the sphere whose directions are indicated by the
arrows. The trajectories described by the black and grey vectors are indicated
by solid and dashed curves, respectively. Evidently, the geometric phase
acquired during the evolution of the Bloch vector in the case $r=\pi/2.1$ (the
solid angle comprehended by the semi-hemisphere above the solid
circumference), is smaller than that for the case $r=\pi/2$ (the solid angle
corresponding to the north hemisphere, equal to $2\pi$).

The solution with $r$ constant means steady population imbalance $\Delta N$,
whereas the relative phase $\phi(t)$, another parameter examined by
experimentalists and necessary to completely define the BS, is a linear
function of time when $\widetilde{\omega}_{a}=\widetilde{\omega}_{b}=0$ or an
oscillating function when $\widetilde{\omega}_{a}=\widetilde{\omega}%
_{b}=\omega_{a0}/4$. Note that the dynamics of the population imbalance and
the relative phase may be followed through the projection of the Bloch vector
trajectory on $z$-axis and $x$-$y$ plane, respectively.

\subsection{On-resonant solution}

As indicated in the Appendix, through the constant of motion $\mathcal{C}%
=\sin\left[  r(t)\right]  \cos\left[  \phi(t)-\delta(t)\right]  $ we obtain
the solution of the characteristic equations (\ref{7}) in the on-resonant
regime where $\Delta(t)=\omega(t)$. All the possible trajectories in the
portrait space $r(t)$ $\times$ $\left(  \phi(t)-\delta(t)\right)  $, are
restrained to the level curves obtained as projection of the surface plotted
in Fig.3, which follows from $\mathcal{C}$.

To better understand the geometric phase acquired by the state vector
$\left\vert \Psi(t)\right\rangle $ in the on-resonant solutions, we consider
two different cases, $\Delta=0$ and $\Delta\neq0$, and analyze its dependence
on the constant $\mathcal{C}$, through the relation%
\begin{align}
\phi_{G}(t)  &  =N\arg\left\{  \cos\left(  r_{0}/2\right)  \cos\left[
r\left(  t\right)  /2\right]  +e^{-i\left[  \phi(t)-\phi_{0}\right]  }%
\sin\left(  r_{0}/2\right)  \sin\left[  r\left(  t\right)  /2\right]  \right\}
\nonumber\\
&  -\frac{N}{2}%
%TCIMACRO{\dint \limits_{t_{0}}^{t}}%
%BeginExpansion
{\displaystyle\int\limits_{t_{0}}^{t}}
%EndExpansion
dt^{\prime}\left\{  \omega(t^{\prime})\left\{  1-\cos\left[  r\left(
t^{\prime}\right)  \right]  \right\}  +\frac{2\mathcal{C}g(t^{\prime}%
)\cos\left[  r\left(  t^{\prime}\right)  \right]  }{1+\cos\left[  r\left(
t^{\prime}\right)  \right]  }\right\}  \text{.} \label{fg2}%
\end{align}

\subsubsection{The case $\Delta=0$}

In Fig.4 we plot the evolution of the geometric phase against $\tau=g_{0}t$,
considering $\widetilde{\omega}_{a}=\widetilde{\omega}_{b}=\widetilde{g}=0$.
The thick solid line on the abscissa axis corresponds to the choice $r_{0}%
=\pi$ and $\phi_{0}=\pi/2$, leading to $\mathcal{C}=0$, under which the
geometric phase is null or undefined as indicated by the open dots over the
abscissa axis. Note that for $r_{0}=\pi$ we get, at $t=0$, an undetermined
equation (\ref{A4b}) for $\phi(t)$. To circumvent such indetermination we
impose on Eq.(\ref{A3}) the constraint $\phi(t)-\delta(t)=(2n+1)\pi/2$ over
any time interval, to determine $\phi(t)$ independently of Eq.(\ref{A4b}).
Since for $\Delta=0$ it follows that $\delta(t)=\delta_{0}$, implying that
$\phi(t)=\delta_{0}+(2n+1)\pi/2$, the geometric phase for the case
$\mathcal{C}=\Delta=0$ simplifies to $\phi_{G}(t)=N\arg\left\{  \cos\left[
\left(  r\left(  t\right)  -r_{0}\right)  /2\right]  \right\}  $ and,
consequently, $\phi_{G}(t)$ is null for $\left\vert r\left(  t\right)
-r_{0}\right\vert \leq\pi$ and undefined for $\left\vert r\left(  t\right)
-r_{0}\right\vert =\pi$. Still in Fig. 4, the solid and dashed lines,
associated to the pairs $(r_{0}$, $\phi_{0})=(\pi/5$, $\pi/4)$ and $\left(
\pi/4\text{, }3\pi/10\right)  $, respectively, correspond to the same constant
$\mathcal{C}\simeq0.41$. These curves exhibits similar behaviors due to the
fact that, with the same constant $\mathcal{C}$, they present the same
trajectory on the portrait space of Fig.3, despite starting from different
initial conditions. The dotted and dashed-dotted lines, associated to the
pairs $\left(  \pi/3\text{, }0\right)  $ and $\left(  \pi/3\text{, }%
\pi\right)  $, and corresponding to the constants $\mathcal{C}\simeq0.87$ and
$\mathcal{C}\simeq-0.87$, respectively, are symmetric around the abscissa axis
$\tau$. Such property of symmetry reflection of the geometric phase $\phi
_{G}\rightarrow-\phi_{G}$, is consequence of the change $\phi_{0}%
\rightarrow\phi_{0}\pm\pi$, implying that $\mathcal{C}\rightarrow-$
$\mathcal{C}$. It is worth noting that the larger the absolute value of
$\mathcal{C}$ the smaller the acquired geometric phase and \textit{vice-versa}%
. \ 

In Fig.5 we plot the evolution of the Bloch vectors coming from the
on-resonant solution with the initial conditions $\left(  \pi\text{, }%
\pi/2\right)  $ and $\left(  \pi/3\text{, }0\right)  $ corresponding to
$\mathcal{C}=0$ and $\mathcal{C}\simeq0.87$, whose initial and final positions
are represented by the black and grey vectors, respectively. As in Fig.2, we
consider the evolution of both vectors during the same time interval $\tau
=\pi$. Through the solid line trajectory described by the black vector, which
oscillates between the north and south poles, it is straightforward to
conclude that the geometric phase is null during the whole time evolution,
except when the vector reaches the north pole, where the geometric phase
becomes undetermined. As the dashed trajectory of the grey vector is not
restricted to a meridian, as in the case $\mathcal{C}=0$, the geometric phase
acquired is evidently non-null.

\subsubsection{The case $\Delta\neq0$}

As we are interested in the dependence of the geometric phase on constant
$\mathcal{C}$ and, now, on the effective frequency of the two-mode condensate
$\Delta(t)=\omega\left(  t\right)  $, we consider all the parameters of the
Hamiltonian being time-independent, $\widetilde{\omega}_{a}=\widetilde{\omega
}_{b}=\widetilde{g}=0$, except for $\delta(t)=\delta_{0}+%
%TCIMACRO{\dint \nolimits_{t_{0}}^{t}}%
%BeginExpansion
{\displaystyle\int\nolimits_{t_{0}}^{t}}
%EndExpansion
\omega(t^{\prime})dt^{\prime}$. In Fig.6 we plot the geometric phase against
$\tau=g_{0}t$ for different initial conditions ($r_{0}$, $\phi_{0}$) and
effective frequencies $\Delta$. As indicated by the thick solid line
associated to the initial conditions $\left(  \pi\text{, }\pi/2\right)  $,
corresponding to $\mathcal{C}=0$, with $\omega_{a}=2\omega_{b}=g_{0}/10$, the
geometric phase is not null, differently from the case $\Delta=0$. The
discontinuity exhibited by this curve follows from the $\arg$ function, whose
characteristic jumps occurs whenever $\tau_{n}=\left[  \left(  2n-1\right)
\pi g_{0}\right]  /2\omega$, $n$ being a positive integer. As indicated by the
solid and dashed lines, the property of symmetry reflection of the geometric
phase in the abscissa axis still follows when changing, simultaneously,
$\phi_{0}\rightarrow\phi_{0}\pm\pi$, and $\omega$ to $-\omega$. In fact, the
solid line corresponds to the initial condition $\left(  \pi/3\text{,
}0\right)  $ with $\mathcal{C}\simeq0.87$ and $\omega_{a}=2\omega_{b}%
=g_{0}/10$, while the dashed line corresponds to $\left(  \pi/3\text{, }%
\pi\right)  $, with $\mathcal{C}\simeq-0.87$ and $\omega_{b}=2\omega_{a}%
=g_{0}/10$. When the change $\phi_{0}\rightarrow\phi_{0}\pm\pi$, are not
followed by that $\omega$ $\rightarrow$ $-\omega$, such a symmetry is not
accomplished as indicated by the dotted line corresponding to the initial
conditions $\left(  \pi/3\text{, }\pi\right)  $ with $\mathcal{C}\simeq-0.87$
and, now, $\omega_{a}=2\omega_{b}=g_{0}/10$.

To visualize the acquisition of the geometric phase we again return to the
evolution of the Bloch vector. As observed from Fig.7, the solid trajectory
described by the vector associated to $\mathcal{C}=0$, whose coincident
initial and final positions are indicated by the black vector, leads to a
finite solid angle and, consequently, a finite geometric phase (during the
time evolution $\tau=\pi$\ considered for both cases presented). The control
of this solid angle may be accomplished through the parameter $\Delta$ --- the
larger $\Delta$ the larger the solid angle and \textit{vice-versa} --- as
demonstrated experimentally through the polarization vector of a photon
undergoing a Mach-Zehder interferometer \cite{Kwiat}. The evolution of the
grey vector, associated to $\mathcal{C}\simeq0.87$, exhibits a dashed
trajectory where the initial and final positions are slightly different. As
the time evolution proceeds, such trajectory leads to a geometric phase which
easily exceed that of the case $\mathcal{C}=0$, as indicated in Fig.6.

The population imbalance $\Delta N$ for the cases $\Delta=0$ and $\Delta\neq0$
is a time-oscillating function strictly dependent on the shape of Rabi
frequency $g(t)$, thus exhibiting a strong connection between the\ dynamics of
population inversion in two-level systems and population imbalance of BECs. In
fact, when written the Hamiltonian (\ref{1}) through the quase-spin operators
in Eqs.(\ref{17}), we obtain, apart from the collision terms, a driven
interaction. On the other hand, the relative phase $\phi(t)$ depends also on
the detuning $\delta(t)$ besides $g(t)$, as shown by Eqs.(\ref{A4}).

\subsection{Off-resonant solution}

To analyze the off-resonant solution, where the detuning $\Delta
(t)=\omega(t)-\varpi$ is controlled by adjusting the parameter $\varpi$, we
impose a constant Rabi frequency $g(t)=g_{0}$ which implies a constant of
motion $\mathcal{C}=\eta\sin\left[  r(t)\right]  \cos\left[  \phi
(t)-\delta(t)\right]  -\cos\left[  r(t)\right]  $. Similarly to the
on-resonant case, all the possible trajectories for the off-resonant $r(t)$
and $\phi(t)$ are restrained to the level curves of the surface following from
$\mathcal{C}$, presented in Fig.8. When $\eta=$ $2g_{0}/\varpi\gg1$ it is
verified that the constant $\mathcal{C}$ reduces to that of the on-resonant
solutions, unless for the multiplicative factor $\eta$, and the surface
presented in Fig.8 also reduces to that of Fig.3. However, for $\eta\ll1$ we
obtain an approximated constant value of $r$. Finally, when $\eta\sim1$, we
obtain the surface whose level curves encapsulate all the possible
trajectories in the portrait space $r(t)$ $\times$ $\phi(t)-\delta(t)$, as
shown in Fig.8.

To study the effect of $\eta$ on the off-resonant geometric phase, given
by\textbf{ }%
\begin{align}
\phi_{G}(t)  &  =N\arg\left\{  \cos\left(  r_{0}/2\right)  \cos\left[
r\left(  t\right)  /2\right]  +e^{-i\left[  \phi(t)-\phi_{0}\right]  }%
\sin\left(  r_{0}/2\right)  \sin\left[  r\left(  t\right)  /2\right]  \right\}
\nonumber\\
&  -\frac{N}{2}%
%TCIMACRO{\dint \limits_{t_{0}}^{t}}%
%BeginExpansion
{\displaystyle\int\limits_{t_{0}}^{t}}
%EndExpansion
dt^{\prime}\left\{  \omega(t^{\prime})\left\{  1-\cos\left[  r\left(
t^{\prime}\right)  \right]  \right\}  +\varpi\cos\left[  r\left(  t^{\prime
}\right)  \right]  \frac{C+\cos\left[  r\left(  t^{\prime}\right)  \right]
}{1+\cos\left[  r\left(  t^{\prime}\right)  \right]  }\right\}  \text{.}
\label{fg3}%
\end{align}
we plot in Fig.9 the evolution of $\phi_{G}(t)$ for different values of $\eta
$, all starting from the point ($\pi/3$, $0)$, with $\omega_{a0}=2\omega
_{b0}=g_{0}/10$ and $\widetilde{\omega}_{a}=\widetilde{\omega}_{b}=0$. The
dotted line, following from the case where $\eta=40$, indicates a similar
behavior to the corresponding case of the on-resonant solution (represented by
the solid line in Fig. 6). For $\eta=0.1$, the solid line shows that the
geometric phase is a strongly oscillating function, a behavior that is better
visualized through the evolution of the corresponding vector in Bloch sphere,
Fig.10. Finally, when $\varpi\sim g_{0}$, as for $\eta=2$, the geometric phase
shows discontinuities, as indicated by the thick solid line, which turns out
to be a signature of the off-resonant solution. We note that the property of
reflection exhibited by the geometric phase coming from the on-resonant
solution is also present here when substituting, simultaneously, $\phi
_{0}\rightarrow\phi_{0}\pm\pi$, $\omega\rightarrow-\omega$, and $\eta
\rightarrow-\eta$.

In Fig.10 we present the evolution of the Bloch vectors in the time interval
$\tau\simeq9\pi/10$, for the cases $\eta=0.1$ and $\eta=2$, both starting from
the common point $\left(  \pi/3\text{, }0\right)  $. The black vector for
$\eta=0.1$, presents a behavior limited to the north hemisphere, described by
the solid trajectory, which exhibits periodic up and down motions on the
parallels, responsible for the oscillations of the geometric phase showed in
Fig.9. The grey vector for $\eta=2$, by its turn, indicates a rather
complicated dashed trajectory which descends to the south hemisphere and goes
back to the north.

Besides depending on the Rabi frequency, as in the on-resonant case, the
population imbalance for the off-resonant solution also depends on the
detuning $\varpi$ which makes its mean value not null. The relative phase
depends on $\delta(t)$, $g(t)$, and the detuning $\varpi$.

\subsection{Time-Dependent Effects on the Geometric Phases}

The behavior of the geometric phase when both, the trap and Rabi frequencies
are TD harmonic functions, as described by Eq.(\ref{27}), is analyzed in
Fig.11, where we plot $\phi_{G}(\tau)$ against $\tau=g_{0}t$, considering the
same initial conditions $(\pi/3$, $0)$ for all the curves. Starting with the
resonant solution with $\Delta=0$, we obtain the dotted curve for the
parameters $\widetilde{\omega}_{a}=\widetilde{\omega}_{b}=0$ and
$\widetilde{g}=\mu=g_{0}$, to be compared with the dotted curve of Fig.4. We
observe that both dotted curves are very close to each other, with the the
increasing rate of the geometric phase being modulated by the TD Rabi
frequency in Fig.11. The solid line corresponds to the on-resonant solution
with $\Delta\neq0$, for the parameters $\omega_{a0}=2\omega_{b0}=g_{0}/10$,
$\widetilde{\omega}_{a}=\widetilde{\omega}_{b}=0$, and $\widetilde{g}%
=\mu=g_{0}$. This curve is to be compared with the solid line in Fig.6,
showing again that the increasing rate of $\phi_{G}$ can be controlled through
the TD Rabi frequency. The dashed line, also corresponding to $\Delta\neq0$,
with $\omega_{a0}=2\omega_{b0}=g_{0}$, $\widetilde{\omega}_{a}=\widetilde
{\omega}_{b}=0$, and $\widetilde{g}=\mu=g_{0}$, shows that the increasing rate
of the geometric phase may also be controlled through the trap frequencies.
Finally, the thick solid line, corresponding to the off-resonant solution,
with $\omega_{a0}=2\omega_{b0}=g_{0}$, $\widetilde{\omega}_{a}=\widetilde
{\omega}_{b}=\chi_{a}=\chi_{b}=g_{0}/2$, $\xi_{a}=0$, $\xi_{b}=-\pi/2$, and
$\widetilde{g}=0$, to be compared with the thick solid line of Fig.9, also
indicates the important role played by the time dependence of the trap
frequency in the geometric phase. We finally observe that, evidently, these TD
effects have direct implications on the population imbalance and relative
phase between the condensates.

\section{Geometric Phases and the External Josephson Effect}

In this section we analyze the two-mode condensates from a different
perspective: as a single atomic specie trapped in a symmetric or asymmetric
double-well potential. As the internal Josephson effect is here substituted by
the tunneling interaction, the laser pumping becomes unnecessary and we impose
that $\Delta=0$, such that $\delta=0$. Moreover, in the external Josephson
effect the interspecies collision rate correspond to a second order correction
compared to the intraspecies collision rates, justifying the assumption that
$\gamma_{ab}\simeq0$ \cite{Milburn1997,Ketterle}.

The Hamiltonian (\ref{1}) applied to this different physical situation leads,
under the above restrictions, to a similar transformed interaction (\ref{4})
and characteristic equations (\ref{7}). Therefore, the different solutions of
the coupled differential equations apply directly to the external Josephson
effect, with the on- and off-resonant processes describing the symmetric
($\omega=0$) and asymmetric ($\omega\neq0$) wells solutions, respectively.

\subsection{Symmetric wells}

>From the above discussion we readily verify that the solutions for $r(t)$ and
$\phi(t)$, coming from the symmetric wells, are give by Eqs.(\ref{A4}) and
(\ref{A4c}) with the constant of motion $\mathcal{C}=\sin\left[  r(t)\right]
\cos\left[  \phi(t)\right]  $. We observe that the TD tunneling rate $g(t)$ is
accomplished by modulating the amplitude of the counter propagating classical
fields that generate the barrier. Similarly to the internal Josephson effect,
all the possible trajectories for $r(t)$ and $\phi(t)$ follow from the level
curves of the surface plotted in Fig.8, assuming $\delta=0$. The same level
curves were obtained, by numerical methods, in Ref.\cite{Balakrishnan}. The
geometric phases acquired by the evolution of the state vector of the BECs are
given by Fig.4 and the Bloch-vector trajectories by Fig.5, both obtained from
the on-resonant solution of the internal Josephson effect with $\delta=0$.

\subsection{Asymmetric wells}

The solutions of the characteristic equations for the asymmetric wells follow
by imposing constant values for the effective frequency $\omega=\varpi$ (since
$\Delta=0$) and the tunneling rate $g=g_{0}$. We thus obtain the solutions
(\ref{A6}) with $\eta$ replaced by $\widetilde{\eta}=2g_{0}/\omega$ and
$\delta=0$. The constant of motion becomes $\mathcal{C}=\widetilde{\eta}%
\cos\left[  \phi(t)\right]  \sin\left[  r(t)\right]  -\cos\left[  r(t)\right]
$. The phase-space portrait $r(t)\times\phi(t)$ is given by Fig.8 (with
$\delta=0$), whose level curves indicate all the possible trajectories for
$r(t)$ and $\phi(t)$. The same level curves were obtained by numerical methods
in Ref.\cite{Balakrishnan}. As an example of the geometric phase acquired by
the evolution of the state vector in the asymmetric wells, we take the dotted
line curve of Fig.9, corresponding to the case $\omega=\varpi=g_{0}/20$. (The
other two curves in Fig.9 do not satisfy the condition $\omega=\varpi$.) The
trajectory of the Bloch vector associated to $\omega=g_{0}/20$ is
approximately give by the dashed curve in Fig.7.

\section{Concluding Remarks}

In the present work we analyze the dynamics of two interacting condensates,
with a full TD Hamiltonian. Starting from the Hamiltonian (\ref{1}) under the
two-mode approximation, an effective interaction (\ref{9}) is established
under the RWA, provided that the polar $r(t)$ and azimuthal $\phi(t)$ angles,
which define a Bloch state, satisfy coupled differential equations. This
procedure enable us to define a detuning $\Delta(t)$ from the Raman resonance
between the atomic transition, together with an effective frequency for the
condensates $\omega\left(  t\right)  $, as in Eq. (\ref{8}). Thus, two
different solutions arise for the differential equations coupling the
parameters\ $r(t)$ and $\phi(t)$, the on-resonant solution, where
$\Delta(t)=\omega\left(  t\right)  $, and the off-resonant solution where
$\Delta(t)=\omega\left(  t\right)  +\varpi$, $\varpi$ being a constant. After
solving analytically the coupled equations for both regimes, we present a
detailed analyzes of the geometric phases acquired by the Bloch state of the
system, also discussing the relative phase and population imbalance.

A main result of our work is the connection between geometric phases and
constant of motions of the interacting condensates which are identified
through the analytical solutions of the coupled differential equations for
$r(t)$ and $\phi(t)$. For each on- or off-resonant solution we assign a
constant of motion which determine the dynamical behavior of the state of the
system and, consequently, its geometric phase. We also note that these
constants of motions follow from level curves in the portrait space which are
also obtained analytically.

To better visualize the time-evolution of the geometric and relative phases
acquired by the state vector, together with population imbalance, we also
analyze the trajectory of the state vector mapped on Bloch sphere. Finally, we
present a brief discussion of the TD effects of the trap and Rabi frequencies
in the geometric phases together with the connection between its evolution in
both cases of internal and external Josephson coupling.

As in this work we studied only the evolution of an initial Bloch state, the
collision parameters were restricted to the global phase of the evolved state,
which remains a Bloch state under the two-mode approximation. Therefore,
collisions terms, which are also assumed as TD parameters, do not play a
decisive role in our present analyzes. It is worth to consider distinct
initial state to analyze the effects of the collisions parameters in the
dynamics of the geometric phase and population imbalance.

\textbf{Acknowledgments}

We wish to express thanks for the support from CNPq and FAPESP, Brazilian agencies.

\section{Appendix}

\subsection{Analytical solutions of the characteristic equations (\ref{7})}

In this Appendix we present some specific solutions of the characteristic
equations (\ref{7}), following a more detailed treatment in \cite{Salomon}. We
investigate two different regimes of the laser field amplification, the
on-resonant and off-resonant regimes, which are defined comparing the
effective frequency of the two-mode condensate, $\omega(t)$, with the detuning
between the laser field and the Raman transition $\Delta(t)$. As mentioned
above, in the on-resonant regime, where $\Delta(t)=\omega\left(  t\right)  $,
the rate of time variation of the laser field equals the effective frequency
of the two-mode condensate. Otherwise, we have the off-resonant regime.

\subsection{On-resonant process}

Defining $\chi(t)\equiv\phi(t)-\delta(t)$, the characteristic equations
(\ref{7}) becomes%

\begin{subequations}
\label{A1}%
\begin{align}
\overset{.}{r}(t)  &  =2g(t)\sin\left[  \chi(t)\right]  \text{,}\label{A1a}\\
\overset{.}{\chi}(t)  &  =\omega(t)-\Delta(t)+2g(t)\cos\left[  \chi(t)\right]
\cot\left[  r(t)\right]  \text{{,}} \label{A1b}%
\end{align}
such that, in the on-resonant regime we are left with the first-order
differential equation
\end{subequations}
\begin{equation}
\frac{dr}{d\chi}=\tan\chi\tan r\mathrm{.} \label{A2}%
\end{equation}
After integrating Eq.(\ref{A2}) we obtain the constant of motion
\begin{equation}
\sin\left[  r(t)\right]  \cos\left[  \phi(t)-\delta(t)\right]  =\mathcal{C},
\label{A3}%
\end{equation}
with $\mathcal{C}$ depending on the initial values $r_{0}$, $\phi_{0}$, and
$\delta_{0}$. Thus, the resonant solutions of Eqs.(\ref{A1}), are given by
\begin{subequations}
\label{A4}%
\begin{align}
\cos\left[  r(t)\right]   &  =\sqrt{1-\mathcal{C}^{2}}\sin\left[
u(t,t_{0})+\arcsin\left(  \frac{\cos r_{0}}{\sqrt{1-\mathcal{C}^{2}}}\right)
\right]  ,\label{A4a}\\
\phi(t)  &  =\delta(t)+\arccos\left\{  \frac{\mathcal{C}}{\sin\left[
r(t)\right]  }\right\}  , \label{A4b}%
\end{align}
where
\end{subequations}
\begin{equation}
u(t,t_{0})=-2\int_{t_{0}}^{t}g(\tau)d\tau\mathrm{.} \label{A4c}%
\end{equation}

\subsection{Off-resonant process}

Considering the off-resonant regime, where $\Delta(t)=\omega(t)-\varpi$,
$\varpi$ being a constant, Eqs. (\ref{A1}) can again be solved by quadrature
as far as we assume the Rabi frequency $g$ to be also a constant $g_{0}$. In
this regime, defining $\eta=$ $2g_{0}/\varpi$, Eq.(\ref{A3}) is replaced by
\begin{equation}
\eta\cos\left[  \phi(t)-\delta(t)\right]  \sin\left[  r(t)\right]
-\cos\left[  r(t)\right]  =\mathcal{C}\mathrm{,} \label{A5}%
\end{equation}
which again depends on the initial values $r_{0}$, $\phi_{0}$, $\delta_{0}$.
The solutions for this case are given by
\begin{subequations}
\label{A6}%
\begin{align}
\cos\left[  r(t)\right]   &  =\frac{\eta\sqrt{1+\eta^{2}-\mathcal{C}^{2}}%
}{1+\eta^{2}}\sin\left\{  -\varpi\sqrt{1+\eta^{2}}\left(  t-t_{0}\right)
\right. \nonumber\\
&  \left.  +\arcsin\left[  \frac{\left(  1+\eta^{2}\right)  \cos
r_{0}+\mathcal{C}}{\eta\sqrt{1+\eta^{2}-\mathcal{C}^{2}}}\right]  \right\}
-\frac{\mathcal{C}}{1+\eta^{2}}\mathrm{,}\label{A6a}\\
\phi(t)  &  =\delta(t)+\arccos\left\{  \frac{\mathcal{C}+\cos\left[
r(t)\right]  }{\eta\sin\left[  r(t)\right]  }\right\}  \mathrm{.} \label{A6b}%
\end{align}
\bigskip

\subsection{A constant solution for $r$}

Another solution for $r(t)$ and $\phi(t)$ arises when we impose that $r$
remains constant in time. Through this solution, given by
\end{subequations}
\begin{align}
r(t)  &  =r_{0},\label{A8a}\\
\phi(t)  &  =\delta(t)+n\pi=\phi_{0}+\int_{t_{0}}^{t}\left[  \omega
(\tau)+2(-)^{n}g\left(  \tau\right)  \cot\left(  r_{0}\right)  \right]
d\tau\text{, } \label{A8b}%
\end{align}
with $r_{0}\neq n\pi$ and $n$ being an integer, the state vector of the system
acquire only relative phase $\phi(t)$. This formal solution for $\phi(t)$ can
be even simplified noting that the physical implementation of this regime
requires, necessarily, that $g\simeq0$. In fact, for the population imbalance
to be null, due to the constant value for $r$, the Rabi frequency must also be
null. therefore, Eq.(\ref{A8b}) simplifies to%
\begin{equation}
\phi(t)=\phi_{0}+\int_{t_{0}}^{t}\omega(\tau)d\tau\text{.} \label{A9}%
\end{equation}

Fig. 1 Absolute value of geometric phase $\phi_{G}(t)$ against the
dimensionless $\tau=\omega_{a0}t$, for $r$ constant, with $\omega_{a0}%
=2\omega_{b0}=62.5\pi$ Hz.

Fig. 2 Evolution of the Bloch vectors coming from constant solutions of $r$ in
the time interval $\tau=4\pi$. The grey and black vectors correspond to the
initial conditions $(\pi/2,0)$ and $(\pi/2.1,0)$, respectively, with
$\omega_{a0}=2\omega_{b0}=62.5\pi$ Hz and $\widetilde{\omega}_{a}%
=\widetilde{\omega}_{b}=0$.

Fig. 3 The portrait space $r(t)$ $\times$ $\phi(t)-\delta(t)$ obtained as
projection of the surface which follows from the on-resonant constant of
motion $\mathcal{C}=\sin\left[  r(t)\right]  \cos\left[  \phi(t)-\delta
(t)\right]  $.

Fig. 4 Evolution of the geometric phase $\phi_{G}(t)$ against $\tau=g_{0}t$,
for on-resonant solutions of the characteristic equations (\ref{7}), with
$\Delta=0$ and $\widetilde{\omega}_{a}=\widetilde{\omega}_{b}=\widetilde{g}=0$.

Fig. 5 Evolution of the Bloch vectors coming from the on-resonant solutions of
Eqs.(\ref{7}) in the time interval $\tau=\pi$. The black and grey vectors
correspond to the initial conditions $\left(  \pi\text{, }\pi/2\right)  $ and
$\left(  \pi/3\text{, }0\right)  $, with $\Delta=0$ and $\widetilde{\omega
}_{a}=\widetilde{\omega}_{b}=\widetilde{g}=0$.

Fig. 6 Evolution of the geometric phase $\phi_{G}(t)$ against $\tau=g_{0}t$,
for on-resonant solutions of the characteristic equations (\ref{7}), with
$\Delta\neq0$ and $\widetilde{\omega}_{a}=\widetilde{\omega}_{b}=\widetilde
{g}=0$.

Fig. 7 Evolution of the Bloch vectors coming from the on-resonant solutions of
Eqs.(\ref{7}) in the time interval $\tau=\pi$. The black and grey vectors
correspond to the initial conditions $\left(  \pi\text{, }\pi/2\right)  $ and
$\left(  \pi/3\text{, }0\right)  $, with $\Delta\neq0$ and $\widetilde{\omega
}_{a}=\widetilde{\omega}_{b}=\widetilde{g}=0$.

Fig. 8 The portrait space $r(t)$ $\times$ $\phi(t)-\delta(t)$ obtained as
projection of the surface which follows from the off-resonant constant of
motion $\mathcal{C}=\eta\sin\left[  r(t)\right]  \cos\left[  \phi
(t)-\delta(t)\right]  -\cos\left[  r(t)\right]  $.

Fig. 9 Evolution of geometric phase $\phi_{G}(t)$ against $\tau=g_{0}t$, for
off-resonant solutions of the characteristic equations (\ref{7}) and different
values of $\eta$, all starting from the point ($\pi/3$, $0)$, with
$\omega_{a0}=2\omega_{b0}=g_{0}/10$ and $\widetilde{\omega}_{a}=\widetilde
{\omega}_{b}=0$.

Fig. 10 Evolution of the Bloch vectors coming from the off-resonant solutions
of Eqs.(\ref{7}) in the time interval $\tau\simeq9\pi/10$. The black and grey
vectors, corresponding to $\eta=0.1$ and $\eta=2$, respectively, both start
from the common point $\left(  \pi/3\text{, }0\right)  $, with $\omega
_{a0}=2\omega_{b0}=g_{0}/10$ and $\widetilde{\omega}_{a}=\widetilde{\omega
}_{b}=0$.

Fig. 11 Evolution of geometric phase $\phi_{G}(\tau)$ against $\tau=g_{0}t$,
for on- and off-resonant solutions of the characteristic equations (\ref{7}),
considering the same initial conditions $(\pi/3$, $0)$ for all the curves.

\bigskip

\bigskip


\begin{thebibliography}{99}                                                                                               %


\bibitem {Peres}A. {Peres}, Phys. Rev. Lett. \textbf{77}, 1413 (1996).

\bibitem {Horodecki}M. Horodecki, P. Horodecki, and R. Horodecki, Phys. Lett.
A \textbf{223}, 1 (1996).

\bibitem {Simon}R. Simon, Phys. Rev. Lett. \textbf{84}, 2726 (2000).

\bibitem {Wootters}W. K. Wootters, Phys. Rev. Lett. \textbf{80}, 2245 (1998).

\bibitem {Werner}G. Giedke, M. M. Wolf, O. Kr\"{u}ger, R. F. Werner, and J. I.
Cirac, Phys. Rev. Lett. \textbf{91}, 107901 (2003).

\bibitem {Latorre}G. Vidal, J. I. Latorre, E. Rico, and A. Kitaev, Phys. Rev.
Lett. \textbf{90}, 227902 (2003).

\bibitem {BCS}J. Bardeen, L. N. Cooper, and J. R. Schriffer, Phys. Rev.
\textbf{108}, 1175 (1957).

\bibitem {Pan05}F. Pan and J. P. Draayer, Phys. Lett. A \textbf{339}, 403 (2005).

\bibitem {Korbicz}J. Hald, J. L. Sorensen, C. Schori, and E. S. Polzik, Rev.
Lett. \textbf{83}, 1319 (1999); J. M. Geremia, J. K. Stockton, and H. Mabuchi,
Science 304, 270 (2004); J. K. Korbicz, J. I. Cirac, and M. Lewenstein, Rev.
Lett. \textbf{95}, 120502 (2005).

\bibitem {Myatt97}C. J. Myatt, E. A. Burt, R. W. Ghrist, E. A. Cornell, and C.
E. Wieman, Phys. Rev. Lett. \textbf{78}, 586 (1997).

\bibitem {Matthews98}M. R. Matthews, D. S. Hall, D. S. Jin, J. R. Ensher, C.
E. Wieman, E. A. Cornell, F. Dalfovo, C. Minniti, and S. Stringari, Phys. Rev.
Lett. \textbf{81}, 243 (1998).

\bibitem {LeggettRMP}A. J. Leggett, Rev. Mod. Phys. \textbf{73}, 307 (2001).

\bibitem {Hall98b}D. S. Hall, M. R. Matthews, C. E. Wieman, and E. A. Cornell,
Phys. Rev. Lett. \textbf{81}, 1543 (1998).

\bibitem {Matthews99}M. R. Matthews, B. P. Anderson, P. C. Haljan, D. S. Hall,
M. J. Holland, J. E. Williams, C. E. Wieman, and E. A. Cornell, Phys. Rev.
Lett. \textbf{83}, 3358 (1999).

\bibitem {Hall98a}D. S. Hall, M. R. Matthews, J. R. Ensher, C. E. Wieman, and
E. A. Cornell, Phys. Rev. Lett. \textbf{81}, 1539 (1998).

\bibitem {Hines03}A. P. Hines, R. H. McKenzie, and G. J. Milburn, Phys. Rev. A
\textbf{67}, 013609 (2003).

\bibitem {Sanz03}L Sanz, R. M. Angelo, and K. Furuya, J. Phys. A: Math. Gen.
\textbf{36}, 9737 (2003).

\bibitem {CiracGBEC98}J. I. Cirac, M. Lewenstein, K. Molmer, and P. Zoller,
Phys. Rev. A \textbf{57}, 1208 (1998).

\bibitem {Gordon99}D. Gordon and C. M. Savage, Phys. Rev. A \textbf{59}, 4623 (1999).

\bibitem {Dunningham01}J. A. Dunningham and K. Burnett, J. Mod. Opt.
\textbf{48}, 1837 (2001).

\bibitem {Vedral}I. Fuentes-Guridi, J. Pachos, S. Bose, V. Vedral, and S.
Choi, Phys. Rev. A \textbf{66}, 022102 (2002).

\bibitem {Chen04}Z.-D. Chen, J.-Q. Liang, S.-Q. Shen, and W.-F. Xie, Phys.
Rev. A \textbf{69}, 023611 (2004).

\bibitem {Milburn}G. L. Salmond, C. A. Holmes, and G. J. Milburn, Phys. Rev. A
\textbf{65}, 033623 (2002).

\bibitem {Baily}S. A. Baily and M. B. Salamon, Phys. Rev. B \textbf{71},
104407 (2005).

\bibitem {Bruno}P. Bruno, V. K. Dugaev, and M. Taillefumier, Phys. Rev. Lett.
\textbf{93}, 096806 (2004).

\bibitem {Kats}Y. Kats, I. Genish, L. Klein, J. W. Reiner, and M. R. Beasley,
Phys. Rev. B \textbf{70}, 180407(R) (2004).

\bibitem {Zanardi}P. Solinas, P. Zanardi, N. Zanghi, and F. Rossi, Phys. Rev.
A \textbf{67}, 052309 (2003).

\bibitem {Berry98}M. V. Berry, Proc. R. Soc. London A \textbf{392}, 45 (1984).

\bibitem {Aharonovl87}Y. Aharonov and J. Anandan, Phys. Rev. Lett.
\textbf{58}, 1593 (1987).

\bibitem {Samuel88}J. Samuel and R. Bhandari, Phys. Rev. Lett. \textbf{60},
2339 (1988).

\bibitem {Tong04}D. M. Tong, E. Sj\"{o}qvist, L. C. Kwek, and C. H. Oh, Phys.
Rev. Lett. \textbf{93}, 080405 (2004).

\bibitem {Piza}J. G. P. de Faria, A. F. R. D. Piza, M. C. Nemes, Europhys.
Lett. \textbf{62}, 782 (2003).

\bibitem {Carollo}A. Carollo, M. F. Santos, and V. Vedral, Phys. Rev. A
\textbf{67}, 063804 (2003).

\bibitem {Duzzioni}E. I. Duzzioni, C. J. Villas-B\^{o}as, S. S. Mizrahi, M. H.
Y. Moussa, and R. M. Serra, Europhys. Lett. \textbf{72}, 21 (2005).

\bibitem {Bliokh}K. Y. Bliokh and Y. P. Bliokh, Annals of Physics
\textbf{319}, 13 (2005).

\bibitem {Milburn1997}G.J. Milburn, J. Corney, E.M. Wright, and D.F. Walls,
Phys. Rev. A \textbf{55}, 4318 (1997).

\bibitem {Villain99}P. Villain and M. Lewenstein, Phys. Rev. A \textbf{59},
2250 (1999).

\bibitem {Vogels}J. M. Vogels, C. C. Tsai, R. S. Freeland, S. J. J. M. F.
Kokkelmans, B. J. Verhaar, and D. J. Heinzen, Phys. Rev. A \textbf{56}, R1067
(1997); J. P. Burke, Jr., J. L. Bohn, B. D. Esry, and C. H. Greene, Phys. Rev.
Lett. \textbf{80}, 2097 (1998).

\bibitem {Smerzi 1997}A. Smerzi, S. Fantoni, S. Giovanazzi, and S.R. Shenoy,
Phys. Rev. Lett. \textbf{79}, 4950 (1997).

\bibitem {Liliana}L. Sanz, M. H. Y. Moussa, K. Furuya, Ann. Phys.
\textbf{321}, 1206 (2006).

\bibitem {Arecchi72}F. T. Arecchi, E. Courtens, R. Gilmore, and H. Thomas,
Phys. Rev. A \textbf{6}, 2211 (1972).

\bibitem {Dowling94}J. P. Dowling, G. S. Agarwal, and W. P. Schleich, Phys.
Rev. A \textbf{49}, 4101 (1994).

\bibitem {Mukunda1993}N. Mukunda and R. Simon, Ann. Phys. \textbf{228}, 205 (1993).

\bibitem {Balakrishnan}R. Balakrishnan and M Mehta, Eur. Phys. J. D
\textbf{33}, 437 (2005).

\bibitem {Albiez2005}M. Albiez, R. Gati, J. Folling, S. Hunsmann, M.
Cristiani, and M.K. Oberthaler, Phys. Rev. Lett. \textbf{95}, 010402 (2005).

\bibitem {Gordon1998}D. Gordon and C.M. Savage, Phys. Rev. A \textbf{58}, 1440 (1998).

\bibitem {Kwiat}M. Ericsson, D. Achilles, J. T. Barreiro, D. Branning, N. A.
Peters, and P. G. Kwiat, Phy. Rev. Lett. \textbf{94}, 050401 (2005).

\bibitem {Polavieja}G. G. Polavieja and E. Sj\"{o}quist, Am. J. Phys.
\textbf{66}, 431 (1998).

\bibitem {Ketterle}M.R. Andrews, C.G. Townsend, H.-J. Miesner, D.S. Durfee,
D.M. Kurn, and W. Ketterle, Science \textbf{275}, 637 (1997).

\bibitem {Salomon}S. S. Mizrahi, M. H. Y. Moussa, and B. Baseia, Int. J. Mod.
Phys. B \textbf{8}, 1563 (1994); C. J. Villas-Boas, F. R. de Paula, R. M.
Serra, and M. H. Y. Moussa, Phys. Rev. A \textbf{68}, 053808 (2003).

\pagebreak\textbf{Figure captions}
\end{thebibliography}
\end{document}